\documentclass[conference]{IEEEtran}
\IEEEoverridecommandlockouts
\usepackage{cite}
\usepackage{amsmath,amssymb,amsfonts}
\usepackage{algorithmic}
\usepackage{graphicx}
\usepackage{textcomp}
\usepackage{xcolor}
\usepackage{tikz}
\usepackage[font={footnotesize}]{subcaption}
\usepackage[font={footnotesize}]{caption}

\usepackage{import}
\usepackage{enumitem}
\usepackage{dirtytalk}

\usepackage{tabularx,booktabs}
\newcolumntype{C}{>{\centering\arraybackslash}X} 
\usepackage[official]{eurosym}

\usepackage{scalerel}
\usepackage{tikz}
\usetikzlibrary{svg.path}

\definecolor{orcidlogocol}{HTML}{A6CE39}
\tikzset{
  orcidlogo/.pic={
    \fill[orcidlogocol] svg{M256,128c0,70.7-57.3,128-128,128C57.3,256,0,198.7,0,128C0,57.3,57.3,0,128,0C198.7,0,256,57.3,256,128z};
    \fill[white] svg{M86.3,186.2H70.9V79.1h15.4v48.4V186.2z}
                 svg{M108.9,79.1h41.6c39.6,0,57,28.3,57,53.6c0,27.5-21.5,53.6-56.8,53.6h-41.8V79.1z M124.3,172.4h24.5c34.9,0,42.9-26.5,42.9-39.7c0-21.5-13.7-39.7-43.7-39.7h-23.7V172.4z}
                 svg{M88.7,56.8c0,5.5-4.5,10.1-10.1,10.1c-5.6,0-10.1-4.6-10.1-10.1c0-5.6,4.5-10.1,10.1-10.1C84.2,46.7,88.7,51.3,88.7,56.8z};
  }
}

\newcommand\orcidicon[1]{\href{https://orcid.org/#1}{\mbox{\scalerel*{
\begin{tikzpicture}[yscale=-1,transform shape]
\pic{orcidlogo};
\end{tikzpicture}
}{|}}}}

\usepackage{hyperref}

\usepackage{graphicx,wrapfig,lipsum}
\def\BibTeX{{\rm B\kern-.05em{\sc i\kern-.025em b}\kern-.08em
    T\kern-.1667em\lower.7ex\hbox{E}\kern-.125emX}}

\usepackage{blindtext}

\usepackage{adjustbox}

\usepackage[mathlines,switch]{lineno}
\makeatletter
\let\old@ps@headings\ps@headings

\let\old@ps@IEEEtitlepagestyle\ps@IEEEtitlepagestyle
\def\confheader#1{%
    \def\ps@headings{%
        \old@ps@headings%
        \def\@oddhead{\strut\hfill#1\hfill\strut}%
        \def\@evenhead{\strut\hfill#1\hfill\strut}%
    }%
    \def\ps@IEEEtitlepagestyle{%
        \old@ps@IEEEtitlepagestyle%
        \def\@oddhead{\strut\hfill#1\hfill\strut}%
        \def\@evenhead{\strut\hfill#1\hfill\strut}%
    }%
    \ps@headings%
}

\makeatother

\confheader{%
    \parbox{20cm}{\textbf{\textit{Accepted to IEEE Belgrade PowerTech, 2023}}}}

\begin{document}

\title{Exploring Operational Flexibility of Active Distribution Networks with Low Observability}

\author{\IEEEauthorblockN{Demetris~Chrysostomou\orcidicon{0000-0002-3564-4629},~\IEEEmembership{Graduate Student Member,~IEEE,}
        Jose~Luis~Rueda~Torres\orcidicon{0000-0001-7288-0228},~\IEEEmembership{Senior Member,~IEEE,}\\
        and~Jochen~Lorenz~Cremer\orcidicon{0000-0001-9284-5083},~\IEEEmembership{Member,~IEEE,}}
\IEEEauthorblockA{\emph{Department of Electrical Sustainable Energy} \\
\emph{Delft University of Technology} \\
Delft, The Netherlands
    \\\{D.Chrysostomou, J.L.RuedaTorres, J.L.Cremer\}@tudelft.nl \vspace{-1em}}
}


\maketitle

\begin{abstract}
Power electronic interfaced devices progressively enable the increasing provision of flexible operational actions in distribution networks. The feasible flexibility these devices can effectively provide requires estimation and quantification so the network operators can plan operations close to real-time. Existing approaches estimating the distribution network flexibility require the full observability of the system, meaning topological and state knowledge. However, the assumption of full observability is unrealistic and represents a barrier to system operators' adaptation. This paper proposes a definition of the distribution network flexibility problem that considers the limited observability in real-time operation. A critical review and assessment of the most prominent approaches are done, based on the proposed definition. This assessment showcases the limitations and benefits of existing approaches for estimating flexibility with low observability. A case study on the CIGRE MV distribution system highlights the drawbacks brought by low observability.
\end{abstract}

\begin{IEEEkeywords}
distribution network flexibility, network observability, network operators \vspace{-0.5em}
\end{IEEEkeywords}

\section{Introduction}
The large-scale integration of renewable energy alters the dynamics of distribution and transmission system operations and aggravates their need for flexibility. In the past, relatively low forecast errors and fluctuations in generated or consumed power could be successfully managed by system operators (SO) by deploying scarce information from their energy management systems. However, the future power system requires more flexibility in active distribution grid management as the system will have more fluctuations and higher uncertainties from the vast increase of distributed generation, prosumers, and storage. Supporting this system's need for higher flexibility, markets for ancillary services may let system members trade their flexibility. From the system operator's perspective, estimating the total available and feasible flexibility in real time can improve grid reliability and reduce economic costs. Moreover, accurately estimating the real-time flexibility and considering network constraints can provide insights into whether SO can reliably respond to disturbances.

Flexibility in power systems is a term used from various directions. These directions include market design \cite{en11040822}, a specific resource type \cite{hekmat2021data}, or the aggregated distribution network (DN) flexibility offered at the TSO-DSO interconnection \cite{kalantar2019characterizing, silva2018challenges, gonzalez2018determination}. This paper focuses on the flexibility of the TSO-DSO interconnections. For the remainder of this document, flexibility will only account for the aggregated DN flexibility offered at the TSO-DSO interconnection. 

\begin{figure}
\centerline{\def\svgwidth{120bp}
    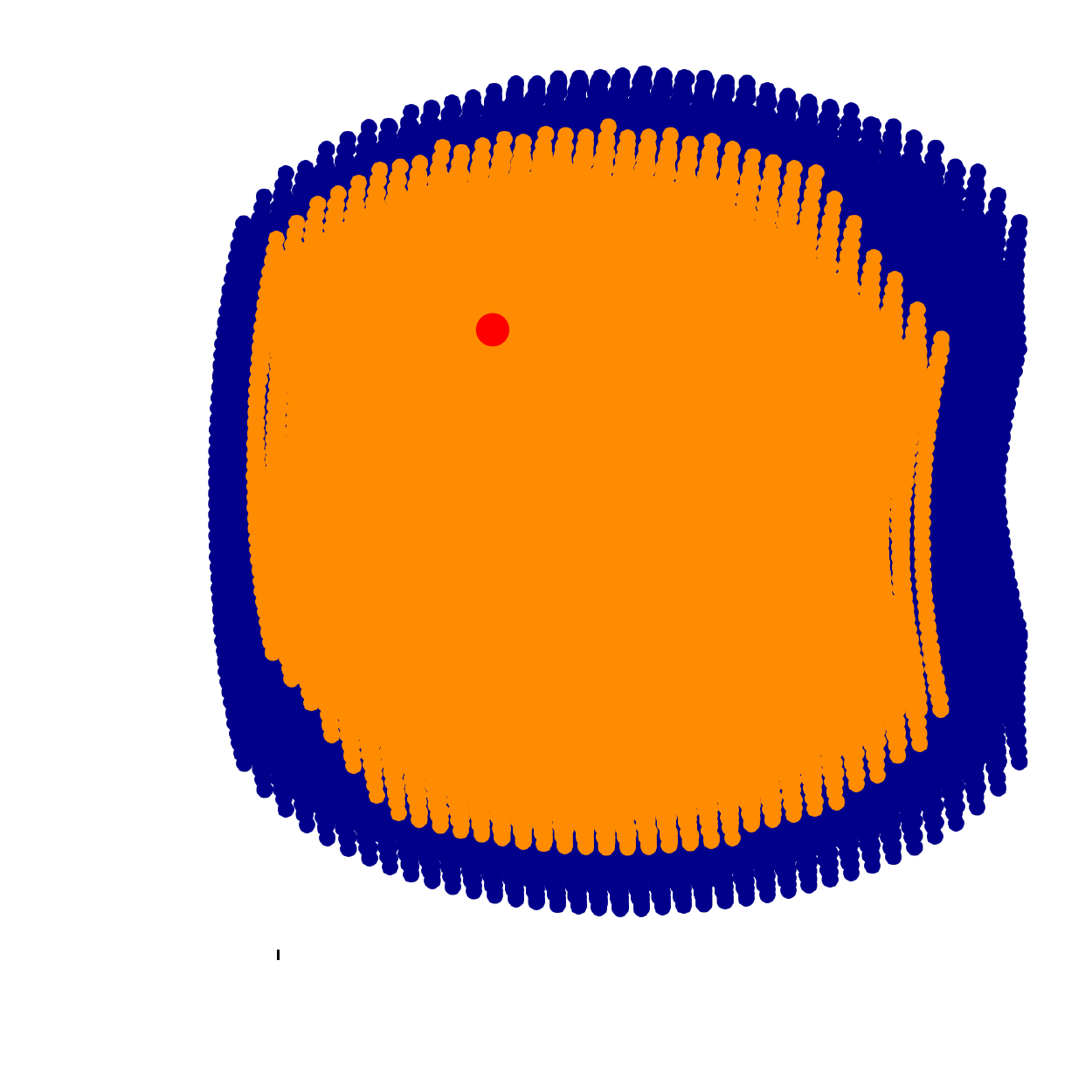 }
    \caption{Example flexibility area of a load and a generator connected to 
 the TSO-DSO interconnection in parallel. Feasible flexibility samples (\protect\includegraphics[height=0.5em]{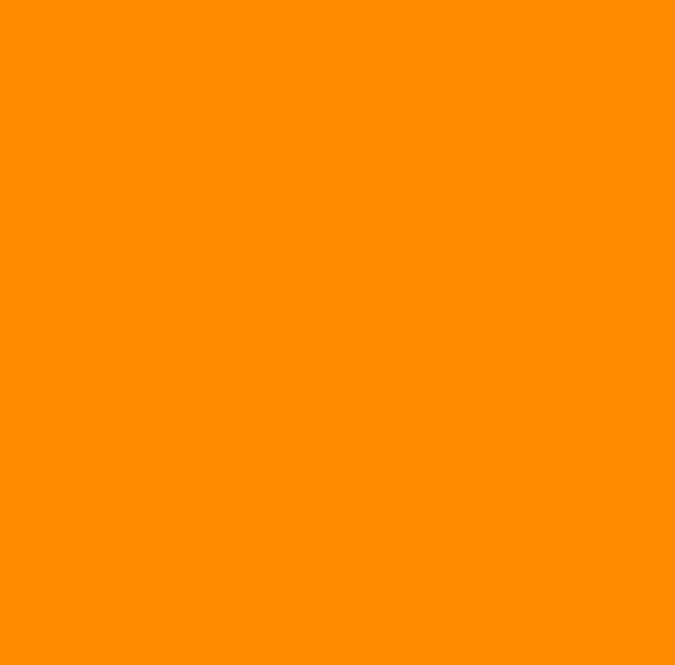}), infeasible flexibility samples (\protect\includegraphics[height=0.5em]{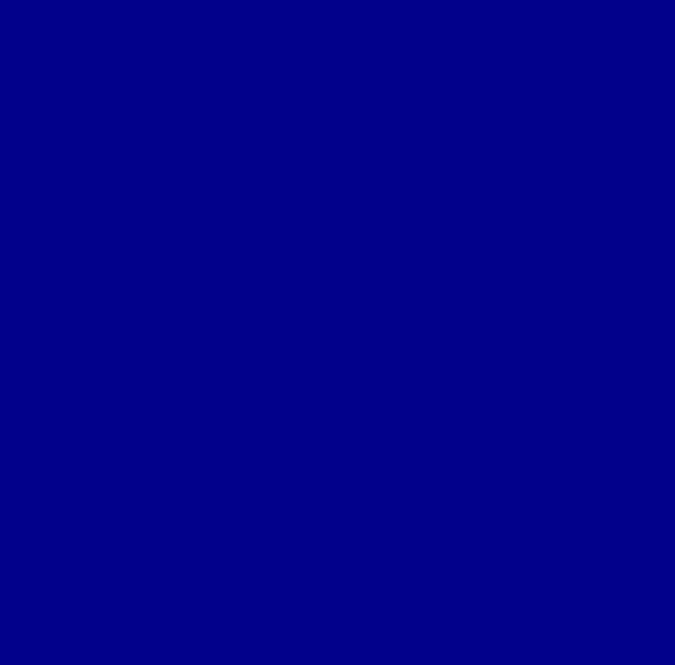}), and initial operating point (\protect\includegraphics[height=0.5em]{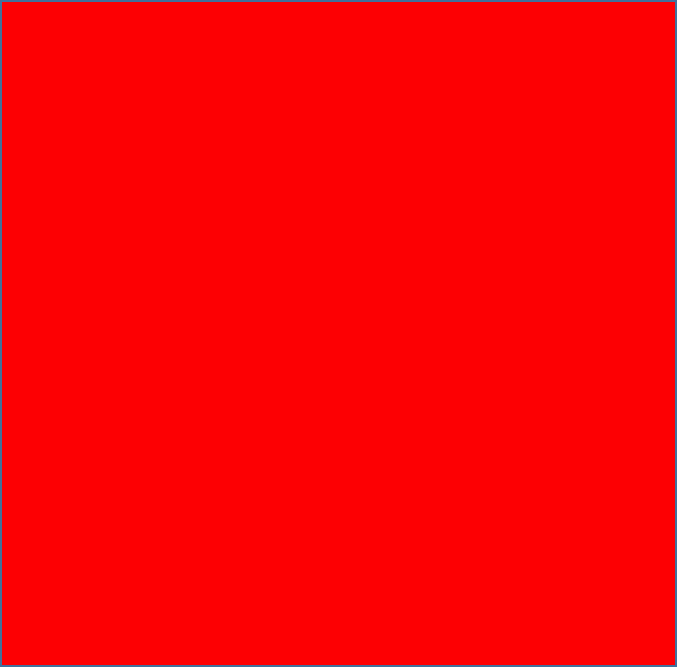}). \vspace{-1em}}
    \label{fig:fa_ex}
\end{figure}

Existing algorithms approach the flexibility estimation problem as range exploration rather than accounting for the population of each flexible point. The objective of existing algorithms that estimate this flexibility is to explore the limits of active and reactive power on the TSO-DSO boundary nodes. Flexible and distributed energy sources providing this flexibility are devices that can alter their operation to help the network avoid technical issues \cite{silva2018challenges}. The flexibility of moving the current operating point to a new one is typically illustrated in a two-dimensional plot. This plot shows the active and reactive power on the TSO-DSO boundary nodes, as shown in Fig. \ref{fig:fa_ex}. The red circle illustrates the initial operating point. The orange area shows the feasible apparent power values in which the operating point can shift using the flexibility of DN sources. The blue area illustrates the values in which the operating point shift through flexibility service providers (FSPs) could cause dissatisfaction with network component voltage or loading. However, the amount of FSP shifts that can reach each feasible point is neither included nor referenced. 

SO should know the available range of flexibility to timely detect in which directions and how much they can shift the TSO-DSO interconnection's operating point to ensure the system's stability and operation. A higher range of flexibility can therefore instigate a bigger space for the SO to move to mitigate any unexpected events. In addition to the flexibility range, SO should know the population of each flexible point. This knowledge can help them anticipate the costs of flexibility, the reliability of the estimated area, and the possibility of delivery of flexibility. Market players should also know the range and population of flexibility since it can allow them to evaluate the need of their offers. Hence, low range and population can impose greater need and monetary compensation for their flexibility. The absence of such insights does not allow the relevant parties to anticipate the number of action sequences reaching each point. 

Assuming the complete system state observability is a common issue of the existing algorithms \cite{bolfek2021analysis, silva2018challenges,  capitanescu2018tso}. The existing algorithms can mainly be categorized as power flow-based and optimal power flow (OPF)-based approaches. Although these approaches have particular strengths and weaknesses, they suffer a similar assumption, the knowledge of network topology and state. However, one issue of DNs is their low observability \cite{7499865}. The algorithm of \cite{kalantar2019characterizing} deals with forecast uncertainty using robust optimization but does not cover topological shifts nor the limited knowledge of non-observable nodes. In addition, when estimating flexibility, the power flow and OPF-based approaches do not represent network dynamics nor time continuity between operating conditions. Accordingly, existing algorithms do not include transients and active network component responses to disturbances.

Definitions of DN flexibility are often non-analytical and designed for a specific purpose or view. An example of these non-analytical definitions can be \textit{"The capability of a distribution network to adjust its active and reactive power flow at the TSO-DSO connection node by using optimization techniques"} \cite{nosair2015flexibility}. Another example is \textit{"The ability of the system components to adjust their operating point, in a timely and harmonized manner, to accommodate expected, as well as unexpected, changes in system operating conditions"} \cite{mohandes2019review}. Further examples can be found in \cite{kalantar2019characterizing, capitanescu2018tso, heleno2015estimation, silva2018challenges} which all define flexibility differently, suited for the application. These definitions do not relate the effects of their assumptions to the objectives of their application. The lack of general definition and the consistency of non-analytical definitions motivated this research to investigate a generalized analytical definition that can later be tailored and clearly interpreted to each use case. The proposed algorithms of existing literature can sometimes only accompany continuous-capability-curve resources \cite{kalantar2019characterizing, capitanescu2018tso}. However, as shown in \cite{silva2018challenges}, discrete variables can produce a different flexibility area than when assumed linear. Therefore, a generalized formulation should be capable to accompany any resource type.

The proposed generalized flexibility definition considers the network locations in which SO measure the active and reactive power flow during operation, and generalizes it to existing flexibility estimation approaches with contributions:
\begin{enumerate}[label=(\roman*)]
    \item an analytical flexibility definition considering the limited observability, flexibility multiplicity, and time continuity.
    \item analysis of the applicability, complexity, and limited observability for possible use cases of this definition.
\end{enumerate}

This paper is organized as follows. Section II describes the proposed generic flexibility definition. Section III discusses the main approaches to estimating flexibility. Section IV provides a case study. Section V summarizes the concluding remarks.

\section{Defining flexibility} \label{definition}
The DN flexibility algorithms rely on theoretical definitions and apply multiple assumptions or simplifications. Simplifications of \cite{capitanescu2018tso, silva2018challenges, churkin2021characterizing, gonzalez2018determination} include ignoring the limited observability in DNs, the network dynamics and transients, the flexibility area time continuity, and the multiplicity of actions.

\subsection{Notions and analytical view of flexibility} \label{notions}

For the proposed definition, the following non-linear equation system is adopted:
\begin{align}
    &\Dot{x}(t) = f(x(t), u(t), v(t), w_x), \label{eq:sys1a}\\
    &\Tilde{x}(t) = r(x(t), \omega_o), \label{eq:sys1b}\\
    &y(t) = h(\Tilde{x}(t), u(t), \Tilde{v}(t), w_y). \label{eq:sys1c}
\end{align}
 $t\in \mathbb{R}_+$ is the time component to include the time-variance of the system, $x(t) \in X \in \mathbb{R}^{2\times n}$ is the time-dependent state matrix representing the active and reactive power at each of the network's $n$ nodes. Furthermore, $u(t) \in U$ is the action matrix representing the active, reactive power shifts for each controllable device. The output $y\in Y\subset \mathbb{R}^2$ is the TSO-DSO interconnection's observable active and reactive power flow. The matrices $w_x,w_o,w_y$ are the system noise, the matrix $v(t) \in \mathbb{R}^{2\times n}$ includes the real and imaginary voltage values per network node, and $\tilde{v}(t)$ are the observable nodal voltages. Furthermore, $r(\cdot)$ is the function limiting the states observable by the SO (i.e., for a completely observable system $r(\cdot) = id(\cdot)$, the identity function). $h(\cdot)$ is the function for flexibility estimation, which uses the actions and observable states to estimate which $y$ are reachable. The system is subjected to inequality technical constraints, e.g., due to allowable min-max limits for voltage magnitudes $c_v^{\textit{min}}\leq g_v(x(t), v(t), u(t))\leq c_v^{\textit{max}}$, branch current flow $g_l(x(t), v(t), u(t))\leq c_l^{\textit{max}}$, and system stability $c_{DVR}^{\textit{min}}\leq g_{DVR}(x(t), v(t), u(t))\leq c_{DVR}^{\textit{max}}$, where DVR is the dynamic variable time response. DVR represents the stability of the network when flexibility is activated, such as dynamic frequency excursions within the time frame of primary frequency control. 

Through this representation, the set of equilibrium points $X_e$, where the power network is steady in the absence of an input and a disturbance, is defined as:
\begin{align}
    X_e = \{x_e| f(x_e(t), 0 , v(t), 0)=0\}.
\end{align}
The definition of the flexibility range for the time window $[\tau_0, \tau_1]$, and the initial state $x(\tau_0)$, can be the set of all $y(\tau_1) \in Y$ in which the system can converge after a disturbance or an action, validating the network constraints. Disturbance can be an event (e.g. a fault) that will cause the FSPs to automatically respond through their controllers, and shift the state from the initial equilibrium point. This range is the orange area of the flexibility example in Fig. \ref{fig:fa_ex}.

If assumed that there are either a finite amount of equilibrium points reachable from $x(\tau_0)$, or $y(t)$ is rounded to $k$ decimals when represented or illustrated, then there is a possibility of multiple $y_1(\tau_1) = y_2(\tau_1)$ reachable from different actions $u_1(\cdot)\neq u_2(\cdot)$. Therefore, a set cannot include all reachable $y(t)$. Nevertheless, a multi-set can be theoretically defined to include the number of actions leading to each $y(t)$. A multiset is an element collection in which some elements may occur more than once \cite{blizard1989multiset}, where $Y$ is the set of distinct elements, and $m: Y\rightarrow \mathbb{Z}_+$ is the multiplicity, which defines the number of times each element of set $Y$ exists. 

\subsection{Proposed definition}
The flexibility of the system described by \eqref{eq:sys1a}--\eqref{eq:sys1c} on the time interval $[\tau_0, \tau_1]$ is the multiset of all $y(\tau_1)\in Y$, and $m_{y(\tau_1)}: Y\rightarrow \mathbb{Z}_+$ whose corresponding $x(\tau_1)$ is an equilibrium point, in which the system can converge after a disturbance or an action, validating the network constraints.

\begin{figure}
\centerline{
\def\svgwidth{140bp}
    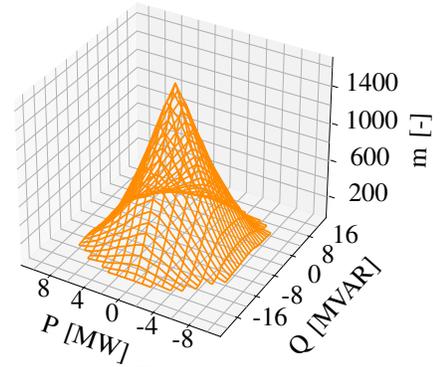 }
    \caption{The flexibility area of Fig. \ref{fig:fa_ex} when considering $m$ the multiplicity.}
    \label{fig:multi}
    \vspace{-0.5em}
\end{figure}

This definition generalizes to the dynamic response of power networks through \eqref{eq:sys1a}, the time variability through the continuous and time-dependent \eqref{eq:sys1a}--\eqref{eq:sys1c}, and low observability through the inclusion of $r(\cdot), \tilde{x}(t)$ in \eqref{eq:sys1b}--\eqref{eq:sys1c}. Considering the disturbance term allows the response of active components, such as grid-forming inverters, to be included as flexibility resources, even without explicitly being shifted through a real-time activation action $u(t)$. The term $u(t)$ is the time-variable shifts of FSPs' power output, i.e., flexibility activation. Finally, this definition includes the range and multiplicity of flexibility by describing flexibility as a multiset instead of a set. Fig. \ref{fig:multi} shows the flexibility when multiplicity $m$ is also accounted for, i.e., the number of different actions $u(t)$ that reach the same feasible $y(t)$ as in \eqref{eq:sys1c}. The projection into the $PQ$ plane results in the range of flexibility as illustrated in Fig. \ref{fig:fa_ex}. The multiplicity of flexible points at the boundaries of the $PQ$-plane projection is significantly lower than the centers. Hence, most actions leading to those points are infeasible due to the network's inequality technical constraints, or reaching those points is possible only through limited action combinations. On the other hand, the highly populated flexible points (with large $m$) near the center suggest multiple options for the SO. 

\subsection{Flexibility estimation objectives and constraints}
The objectives for the estimation of the proposed multiset described flexibility are
\begin{align}
    &\max \left(range(h \circ r)_{[\tau_0, \tau_1]}\right), \label{eq:range} \\
    &\max \left(m(y_i)\right)\quad \forall y_i \in range(h \circ r)_{[\tau_0, \tau_1]}, \label{eq:multi}
\end{align}
here $range(\cdot)$ is the flexibility range described in Sec. \ref{notions}. $h \circ r$ is the composition of flexibility estimation function $h$ and observable state limitation function $r$ \cite{lygeros2010lecture}. The first objective \eqref{eq:range} explores all possible active and reactive power shifts $y$ reachable from an initial starting time $\tau_0$ until the time $\tau_1$. The second objective \eqref{eq:multi} explores the multiplicity $m$ of each reachable shift. Thus, the number of action combinations that exist for each reachable shift. Considering $h(\cdot)$ and $r(\cdot)$ in the objectives highlights the influence of the flexibility estimation algorithm and the observable states on the result. Accordingly, improving the flexibility estimation could be through upgrading the estimation algorithms $h(\cdot)$,  upgrading the network observability levels $r(\cdot)$, or both.

As shown in the generalized flexibility definition, the constraints accompanying the network are nodal under and over-voltage $c_v^{\textit{min}}, c_v^{\textit{max}}$, the branch loading limitations $c_l^{\textit{max}}$, and the minimum and maximum system dynamic variable time response $c_{DVR}^{\textit{min}}, c_{DVR}^{\textit{max}}$. However, including additional constraints based on the needs of the flexibility estimating authority is feasible. As an example, a system operator who has a limitation on funds for flexibility activation can add a constraint $g_{\euro{}}(u)\leq c_\euro{}$. This additional constraint depends on the activation actions, which form costs depending on the shift amount, duration, and type of FSP.


\section{Flexibility estimation for steady-state performance} \label{sec:approach}

\subsection{Power flow-based approaches} \label{sec:pf-based}
Power flow-based approaches, starting from observable networks sample various shifts in the controllable flexibility sources. Using these shifts, they solve power flows to detect whether the new operating points would meet the constraints or not \cite{heleno2015estimation, gonzalez2018determination}. Termination conditions for such algorithms can be the number of power flows run or time spent. Hence, the overall methodology followed by such algorithms has the following four steps:
\begin{enumerate}
    \item Identify the shift capabilities $U$ of each FSP and the initial operating state $x$ of all network nodes.
    \item Sample an operation shift for each FSP within the identified shift capabilities $u \in U$.
    \item Run power flow to obtain the new TSO-DSO node's operating state $y$ and check if it is feasible regarding network constraints.
    \item Termination condition reached?
    \begin{itemize}
        \item If yes, plot flexibility area $Y$.
        \item If not, starting from the initial operating state, sample a different shift per FSP and go to step $3$.
    \end{itemize}
\end{enumerate}
This methodology does not include the \eqref{eq:sys1b} since step $3$ runs the power flow in the complete network, thus taking $x$ as an input. The dynamics of \eqref{eq:sys1a} are ignored since these approaches are for steady state, and output $Y$ does not include multiplicity.

The advantages of power flow-based algorithms include no need for linearization nor convexification of the objectives \eqref{eq:range}--\eqref{eq:multi}. Therefore, non-linear variables (e.g., on-load tap changers (OLTC)) do not affect their performance and can result in non-convex flexibility areas. Limitations of these algorithms include being significantly slower than OPF-based approaches \cite{silva2018challenges}, and their performance's dependence on the sampling distributions used for the FSP actions, i.e., which actions $u$ are sampled and passed to \eqref{eq:sys1c} to find feasible points $y$.

\subsection{Optimal power flow-based approaches}
OPF-based approaches use multi-objective optimization (MOO) to compute the flexibility areas. In general terms, the four objectives that can theoretically be pursued are:
\begin{align}
 &\mathrm{min} \left( P_{TSO- DSO}\right), \label{eq:obj1}\\ 
 &\mathrm{max}\left( P_{TSO- DSO}\right), \label{eq:obj2}\\ 
 &\mathrm{min}\left( Q_{TSO- DSO}\right), \label{eq:obj3}\\
 &\mathrm{max} \left(Q_{TSO- DSO}\right) \label{eq:obj4},  
\end{align}

where $P_{TSO- DSO}, Q_{TSO- DSO}$ are the active and reactive power flowing from the TSO to the DSO, respectively. The approaches adapt the OPF constraints to include flexibility source operating point shifts.

Objectives \eqref{eq:obj1}--\eqref{eq:obj2} and \eqref{eq:obj3}--\eqref{eq:obj4} conflict with each other. Thus, these multi-objective optimization problems cannot combine all objectives in one. Therefore, OPF-based algorithms solve four MOOs for each flexibility area $\mathrm{min}$$ \left(P_{TSO- DSO}\right)$ and $\mathrm{min}$$ \left(Q_{TSO- DSO}\right)$, $\mathrm{min}$$ \left(P_{TSO- DSO}\right)$ and $\mathrm{max}$$ \left(Q_{TSO- DSO}\right)$, $\mathrm{max}$$ \left(P_{TSO- DSO}\right)$ and $\mathrm{min}$$ \left(Q_{TSO- DSO}\right)$, and $\mathrm{max}$$ \left(P_{TSO- DSO}\right)$ and $\mathrm{max}$$ \left(Q_{TSO- DSO}\right)$. These MOOs are solved with approaches such as the $\epsilon$ constraint method \cite{capitanescu2018tso, churkin2021characterizing}, weighted sum method \cite{silva2018challenges}, and radial reconstruction-based method \cite{kalantar2019characterizing, savvopoulos2021contribution}.

The difference between $\varepsilon$-constrained and weighted sum-based approaches is how they combine objectives. The $\varepsilon$-constrained considers only one of the objectives to be optimized and considers the other objectives as inequality constraints greater than $\epsilon$. The $\epsilon$ is adjusted to approach different limits of the flexibility area. The weighted sum approach optimizes a single objective, the weighted summation of each objective of the initial MOO problem. The weights are adjusted to approach different flexibility areas. These two optimization-based approaches have the limitation that their resulting flexibility areas are convex hulls, which might not represent the actual case as explained in \cite{silva2018challenges}. These approaches either perform linearization of non-linear control variables (OLTC) \cite{silva2018challenges, capitanescu2018tso} or do not include non-linear control variables \cite{churkin2021characterizing}. Radial reconstruction iteratively considers an angle $\theta$ of active and reactive power proportions, and it finds the active and reactive power limits along the line of $\theta$ \cite{kalantar2019characterizing}. Radial reconstruction-based approaches include both convex optimization \cite{kalantar2019characterizing} and non-convex optimization \cite{savvopoulos2021contribution}. 

The advantages of OPF-based approaches are their speed and independency with FSP shift sampling distributions. A limitation of the above algorithms is not dealing with disjoint flexibility areas, which can be the case when FSPs include discrete variables \cite{silva2018challenges}. OPF-based approaches ignore the dynamics of \eqref{eq:sys1a} and assess the steady-state performance. The OPF considers the complete network, thus ignoring \eqref{eq:sys1b}. The output of these approaches aims only at objective \eqref{eq:range}.

\subsection{Assessing the existing approaches with the proposed definition}

The approaches analyzed in the previous two subsections can be characterized based on the proposed definition in terms of low observability, multiplicity, and time continuity. The two approaches based on power flow and OPF need the initial network state $x(t)$ to explore the boundaries of the area. Therefore, the $h(\cdot)$ of \eqref{eq:sys1c} takes $x(t)$ as an input, and the low observability of \eqref{eq:sys1b} is neglected. Thus, the objectives \eqref{eq:range} and \eqref{eq:multi} are altered into:
\begin{align}
    &\max \left(range(h)_{[\tau_0, \tau_1]}\right), \label{eq:range2} \\
    &\max \left(m(y_i)\right)\quad \forall y_i \in range(h)_{[\tau_0, \tau_1]}. \label{eq:multi2}
\end{align}
Such assumption could be strong as analyzed in the case study of Sec. \ref{case study}, where limited observability compromises determining the flexibility area. 

Regarding the objectives, OPF-based approaches' objectives \eqref{eq:obj1}--\eqref{eq:obj4} are aligned with the proposed objective of \eqref{eq:range} since pushing the limits of feasible $P_{TSO- DSO}, Q_{TSO- DSO}$ maximizes the set of feasible $y$, and the  range of $h$. However, OPF-based approaches do not deal with \eqref{eq:multi} since \eqref{eq:obj1}--\eqref{eq:obj4} do not include multiplicity. The power flow-based approaches can deal with the objectives \eqref{eq:range2} and \eqref{eq:multi2} as they can sample actions leading to a similar flexibility point. However, power flow-based approaches sample actions using predefined distributions. Hence, the multiplicity reported by power-flow-based approaches can be faulty and biased toward these distributions. In addition, multiplicity results are not visualized nor referenced within the found literature.

The two approaches do not perform dynamic simulations in terms of state continuity and transients. The approaches also do not account for the response that inverter-based generators can have to external disturbances. Thus, the continuous state dynamics are discretized, and the system of \eqref{eq:sys1a}--\eqref{eq:sys1c} becomes:
\begin{align}
    &x(\tau+\Delta \tau) = \hat{f}(x(\tau), u(\tau), v(\tau), \omega_x),\\
    &y(\tau) = h(x(\tau), u(\tau), v(\tau), \omega_y),
\end{align}
Furthermore, the action or shift capabilities of flexibility service providers are typically assumed not to be constrained by resource time characteristics. Hence, $\Delta \tau$ is assumed large enough that all machines can change their outputs to their limits. Thus, $u(\tau)$ is sometimes replaced by $u$.

\section{Case study} \label{case study}
\begin{figure}
    \centering
    \includegraphics[width=0.39\textwidth]{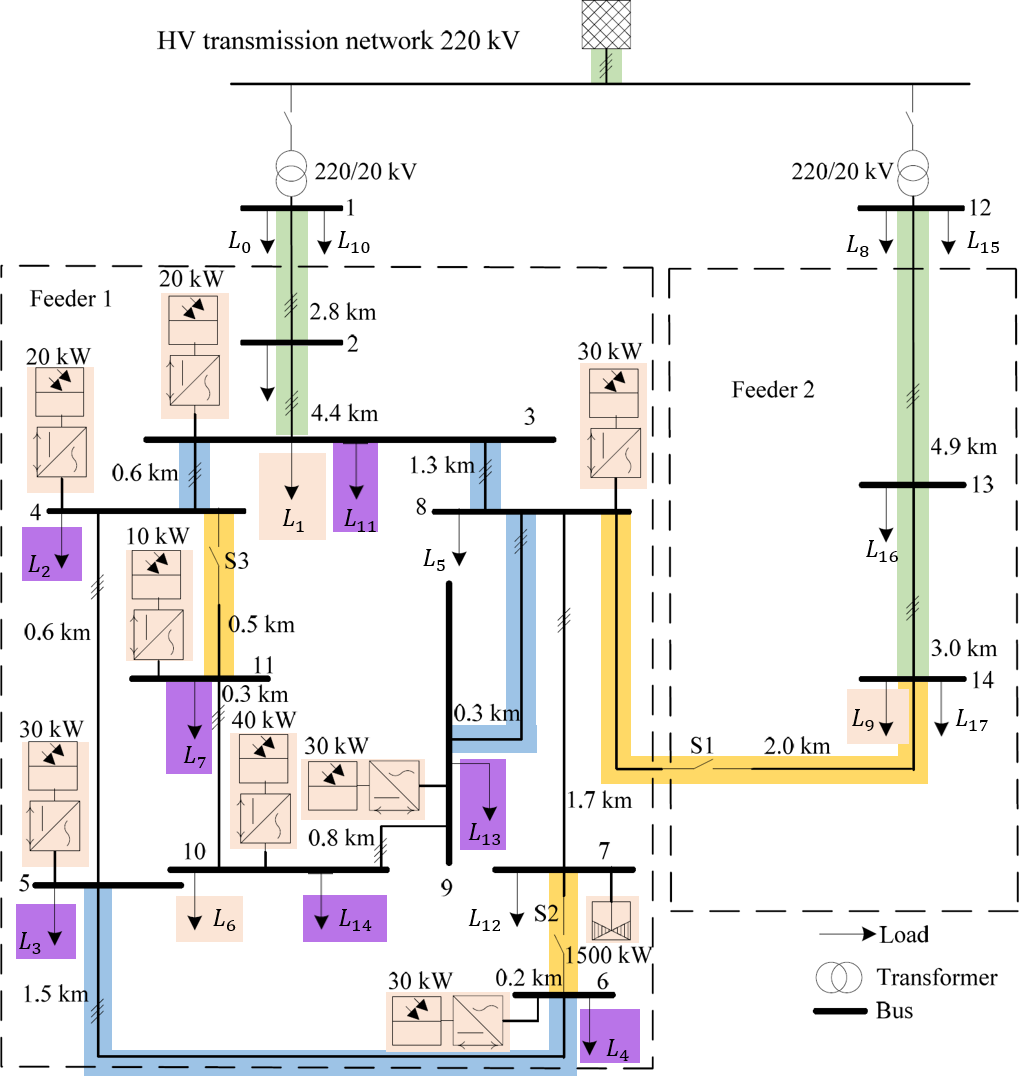}
    \caption{CIGRE MV DN with solar photovoltaic and Wind DER with modifications.Observable lines (\protect\includegraphics[height=0.5em]{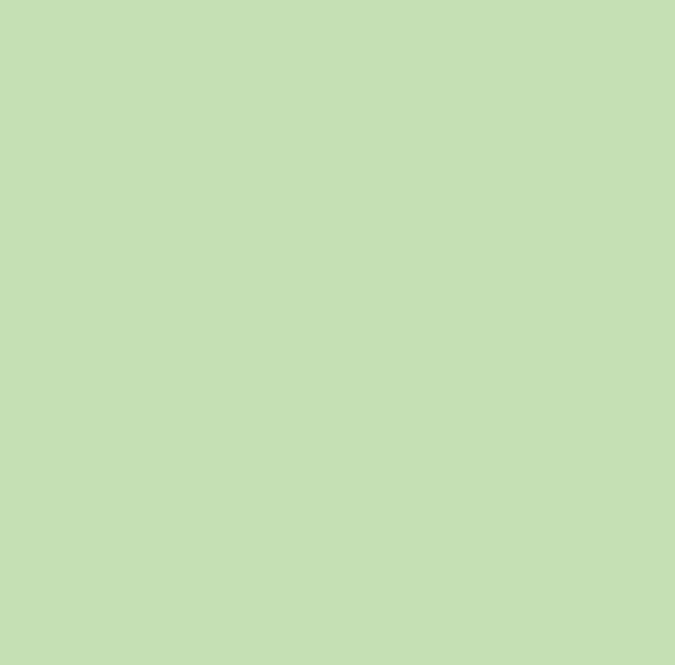}). FSPs (\protect\includegraphics[height=0.5em]{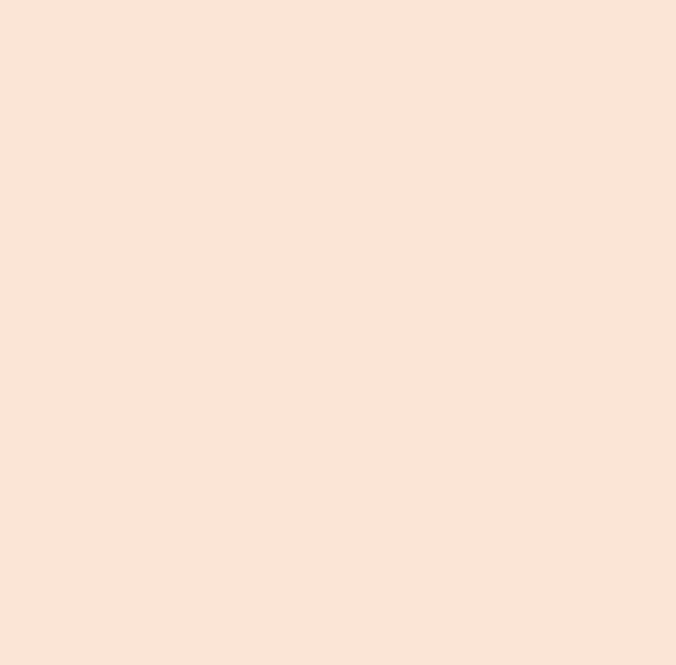}). TSS altered lines (\protect\includegraphics[height=0.5em]{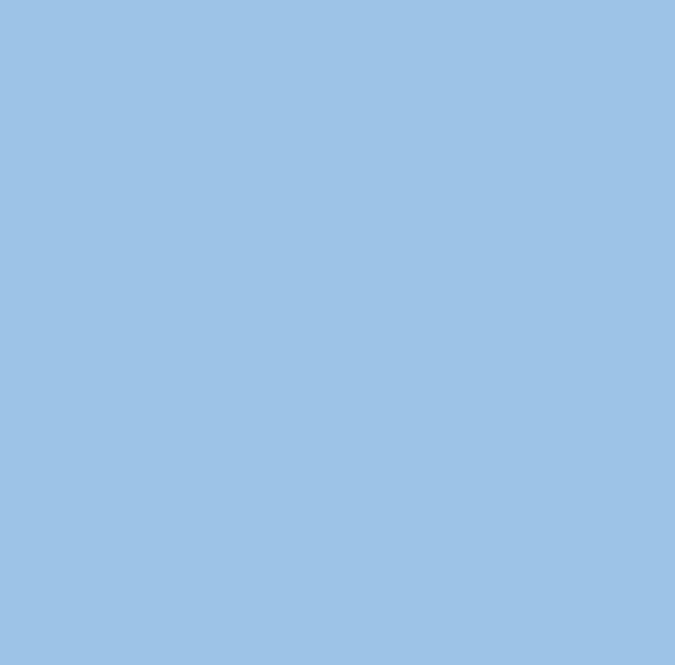}) and switches (\protect\includegraphics[height=0.5em]{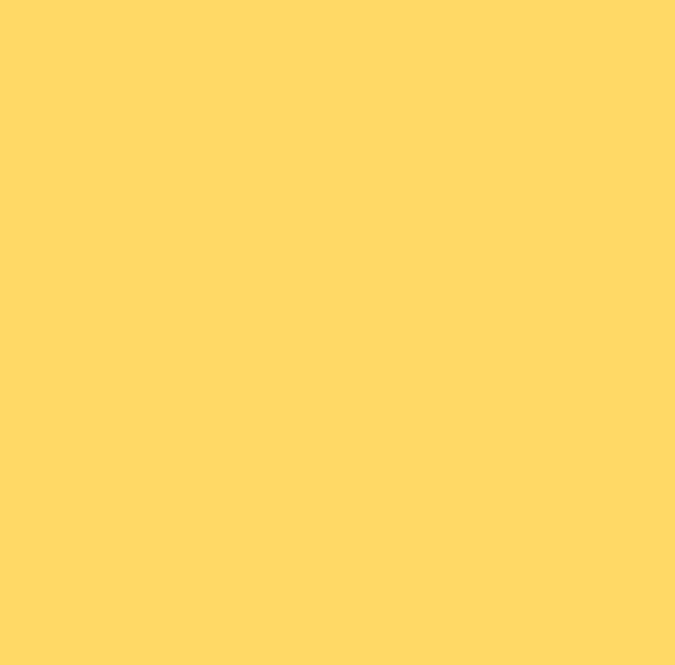}). USS/TSS altered loads (\protect\includegraphics[height=0.5em]{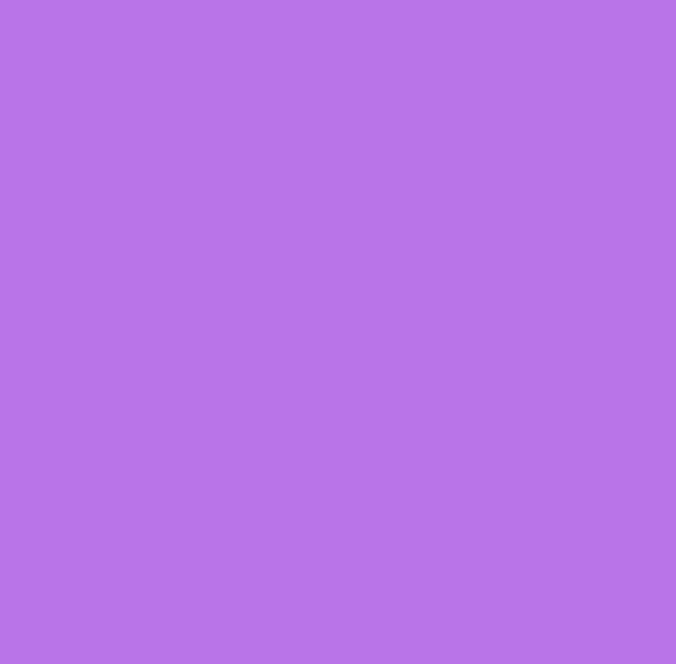}).}
    \label{fig:cigre}
    \vspace{-1em}
\end{figure}

\begin{table*}
\caption{Observable component power difference per scenario} \label{tab:ERR}
\begin{tabularx}{\textwidth}{@{} l *{25}{C} c @{}}
\toprule
\multicolumn{1}{l|}{\textbf{ }}&\multicolumn{2}{c}{\textbf{TSS1}}&\multicolumn{2}{c}{\textbf{TSS2}}&\multicolumn{2}{c}{\textbf{TSS3}}&\multicolumn{2}{c}{\textbf{USS1}}&\multicolumn{2}{c}{\textbf{USS2}}\\
\midrule
\multicolumn{1}{l|}{\textbf{Component}} & $\Delta P[\%]$ & $\Delta Q[\%]$ & $\Delta P[\%]$ & $\Delta Q[\%]$ & $\Delta P[\%]$ & $\Delta Q[\%]$ &  $\Delta P[\%]$ & $\Delta Q[\%]$& $\Delta P[\%]$ & $\Delta Q[\%]$\\
\midrule
\multicolumn{1}{l|}{Line 0} & 0.03 &	-0.17 & -0.03 &	0.02 &	0.03 & 0.47 & -0.05 & -0.02 &	-0.03 & 0.02\\
\multicolumn{1}{l|}{Line 1} & 0.03 &	-0.17 &	-0.03 &	0.02 &	0.02 &	0.46 &	-0.05 &	-0.01 &	-0.03	& 0.02 \\
\multicolumn{1}{l|}{Line 10} & 0.00 & 	0.00 &	0.00 &	0.00	& 0.06 &	0.02 & 0.00 & 0.00 & 0.00 &	0.00\\
\multicolumn{1}{l|}{Line 11} & 0.00 &	0.00 &	0.00 & 0.00 &	0.06 &	0.03 & 0.00 & 0.00 & 0.00 &	0.00 \\
\multicolumn{1}{l|}{External Grid} & 0.00 &	-0.01 &	0.00 &	0.00 &	0.00 &	0.04 &	0.00 &	0.00 &	0.00 &	0.00\\
\bottomrule
\end{tabularx}
\vspace{-1em}
\end{table*}

This case study investigates the observability of DNs when estimating flexibility for steady-state performance. CIGRE medium voltage DN is used with limited real-time observable lines to investigate the flexibility area's sensitivity. In all scenarios, FSPs offer similar flexibility and the observable lines measure approximately similar values. The implementation in Python uses Pandapower's \cite{pandapower.2018} version of the CIGRE medium voltage (MV) DN with solar photovoltaic and wind-distributed energy resources (DER). The network representation in Fig. \ref{fig:cigre}, is modified from \cite{cigre_net} to show the network's loads as obtained from Pandapower \cite{pandapower.2018}. The considered flexible devices are all the DER and the loads $L1, L6, L9$. The real-time observable lines are assumed to be longer than $2.5 km$, thus those connecting the buses $1\textrm{--}2, 12\textrm{--}13, 2\textrm{--}3$, and $13\textrm{--}14$ and the TSO-DSO interconnection line. The low observability scenarios are:
\begin{enumerate}
    \item \textbf{Topological and state shifts (TSS):} A subset of the unobservable lines are re-connected or disconnected. A few unobservable, non-flexible components have different outputs to reduce the differences in the power flowing on the observable lines.
    \item \textbf{Unobservable state initial condition shifts (USS):} The initial power level of some unobservable, non-flexible loads is different from the unaltered network's.
\end{enumerate}
The study case includes three TSS and two USS scenarios. The first TSS scenario reconnects the lines between buses, $4\textrm{--}11$ and $6\textrm{--}7$, disconnects the ones between $3\textrm{--}4$ and $8\textrm{--}9$, and alters the load's $L11$ initial consumption. The second TSS scenario reconnects the same lines and alters the load $L11$ but disconnects the lines between buses $5\textrm{--}6$ and $8\textrm{--}9$ instead. The third TSS scenario reconnects all yellow lines of Fig. \ref{fig:cigre}, disconnects the lines between buses $3\textrm{--}4$ and $3\textrm{--}8$, and changes the initial loads' $L11$ and $L7$ consumption. The two USS scenarios change the initial consumption of the loads $L2,L3,L4,L7,L13$, and $L14$. 
The flexibility area estimation algorithm uses the power flow-based algorithm described by the four steps in Sec, \ref{sec:approach}. The termination condition is set to $100,000$ power flows for each scenario. For the algorithm's step $1$, each scenario has different operating points in unobservable nodes. In step $2$, the algorithm applies operation shifts ($u$) for each FSP. These operation shifts are the same for each scenario, as all scenarios use the same FSPs. After each iteration on step 3 of the algorithm, the tuple $(y, u, x, v)$ is stored. Based on $x,v$, and the limitations $c_v^l=0.95 p.u., c_v^h=1.05p.u., c_l^h=100\%$ the algorithm categorizes the new TSO-DSO interconnection's operating point ($y$) as feasible or infeasible. The DVR constraints are not included since the study case concerns steady-state performance. The high computational burden of the implementation is a concern of focus in future work. The algorithm moves to a subsequent sample if the power flow does not converge. The software and results of the study case are available at \cite{Code}.

\begin{figure}
    \vspace{-0.5em}
     \begin{subfigure}[b]{0.155\textwidth}
         \centerline{\def\svgwidth{80bp}
        \fontsize{7pt}{9pt}\selectfont 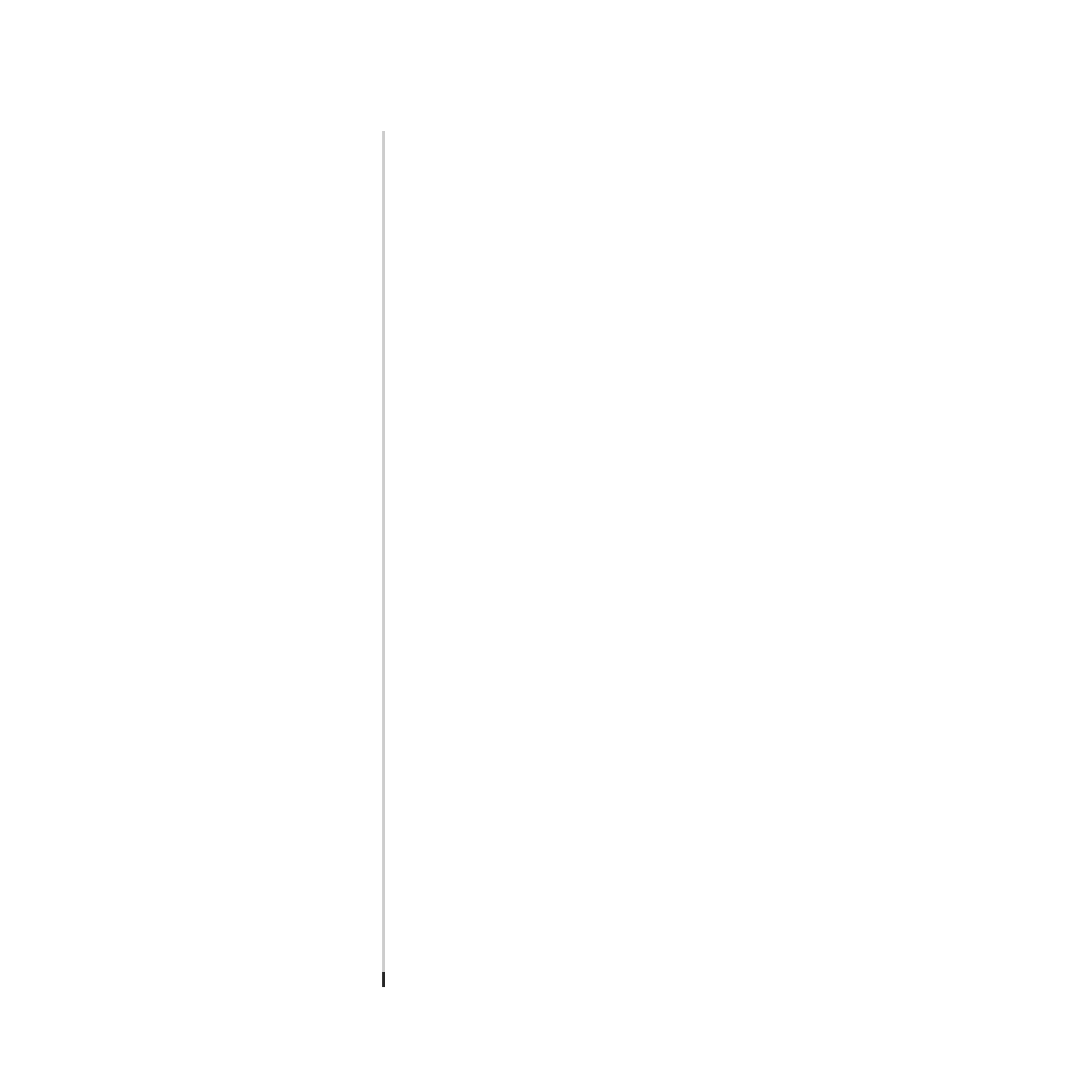}
         \caption{TSS1}
         \label{fig:TSS1}
     \end{subfigure}
          \hfill
     \begin{subfigure}[b]{0.155\textwidth}
         \centerline{\def\svgwidth{80bp}
         \fontsize{7pt}{9pt}\selectfont
         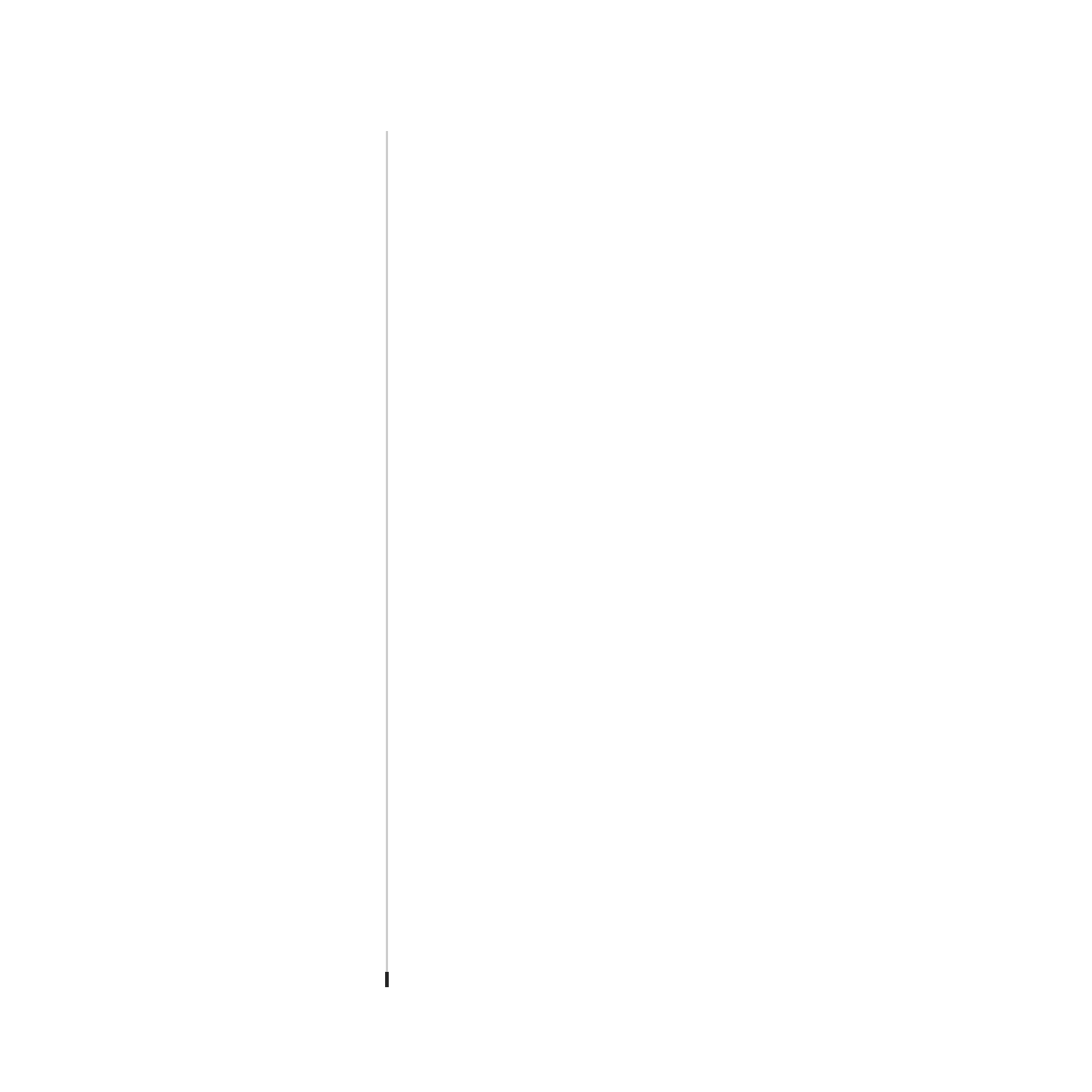}
         \caption{TSS2}
         \label{fig:TSS2}
     \end{subfigure}
          \hfill
     \begin{subfigure}[b]{0.155\textwidth}
         \centerline{\def\svgwidth{80bp}
         \fontsize{7pt}{9pt}\selectfont
         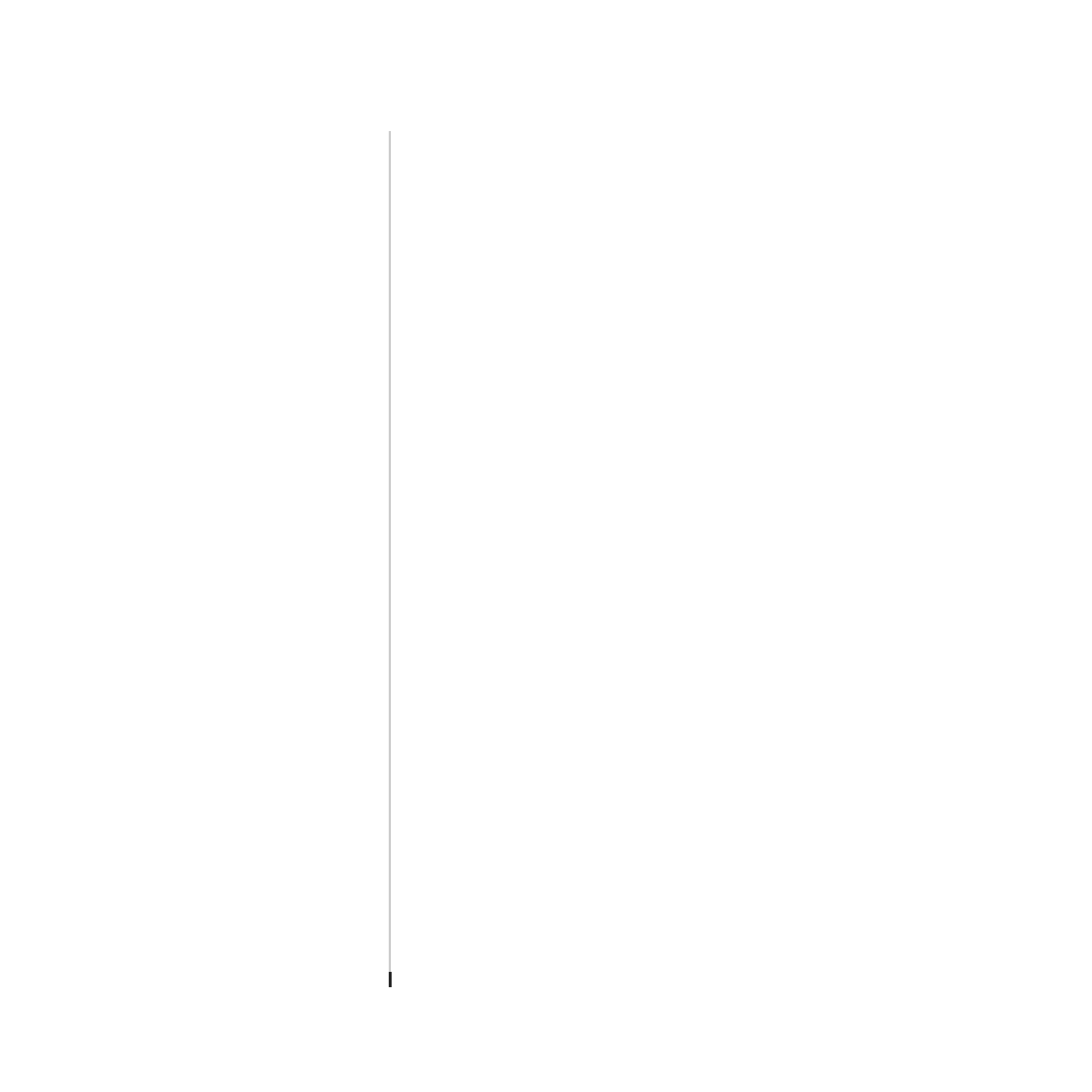}
         \caption{TSS3}
         \label{fig:TSS3}
     \end{subfigure}
          \hfill
     \begin{subfigure}[b]{0.155\textwidth}
         \centerline{\def\svgwidth{80bp}
         \fontsize{7pt}{9pt}\selectfont
         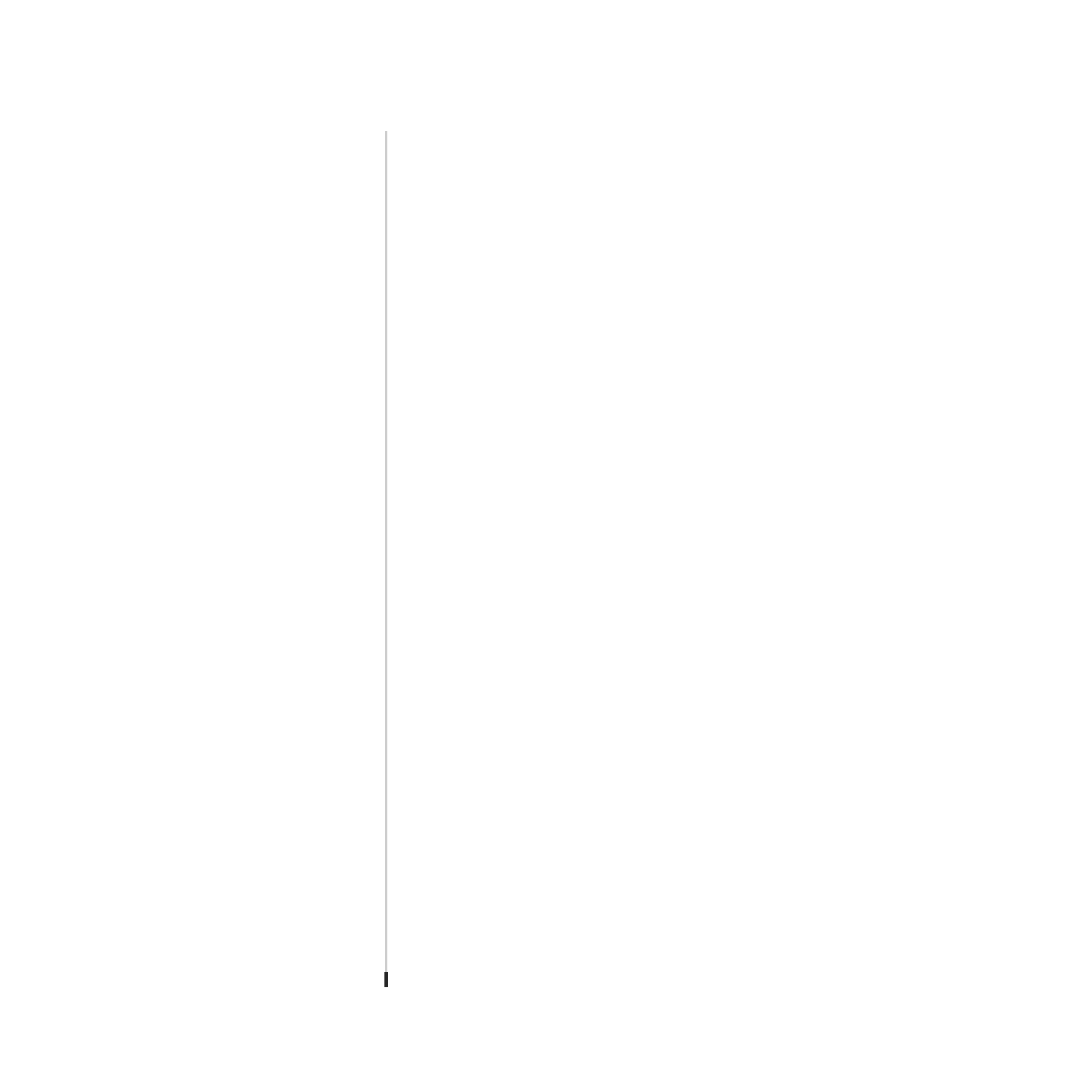}
         \caption{USS1}
         \label{fig:TSS3}
     \end{subfigure}
     \hfill
     \begin{subfigure}[b]{0.155\textwidth}
         \centerline{\def\svgwidth{80bp}
         \fontsize{7pt}{9pt}\selectfont
         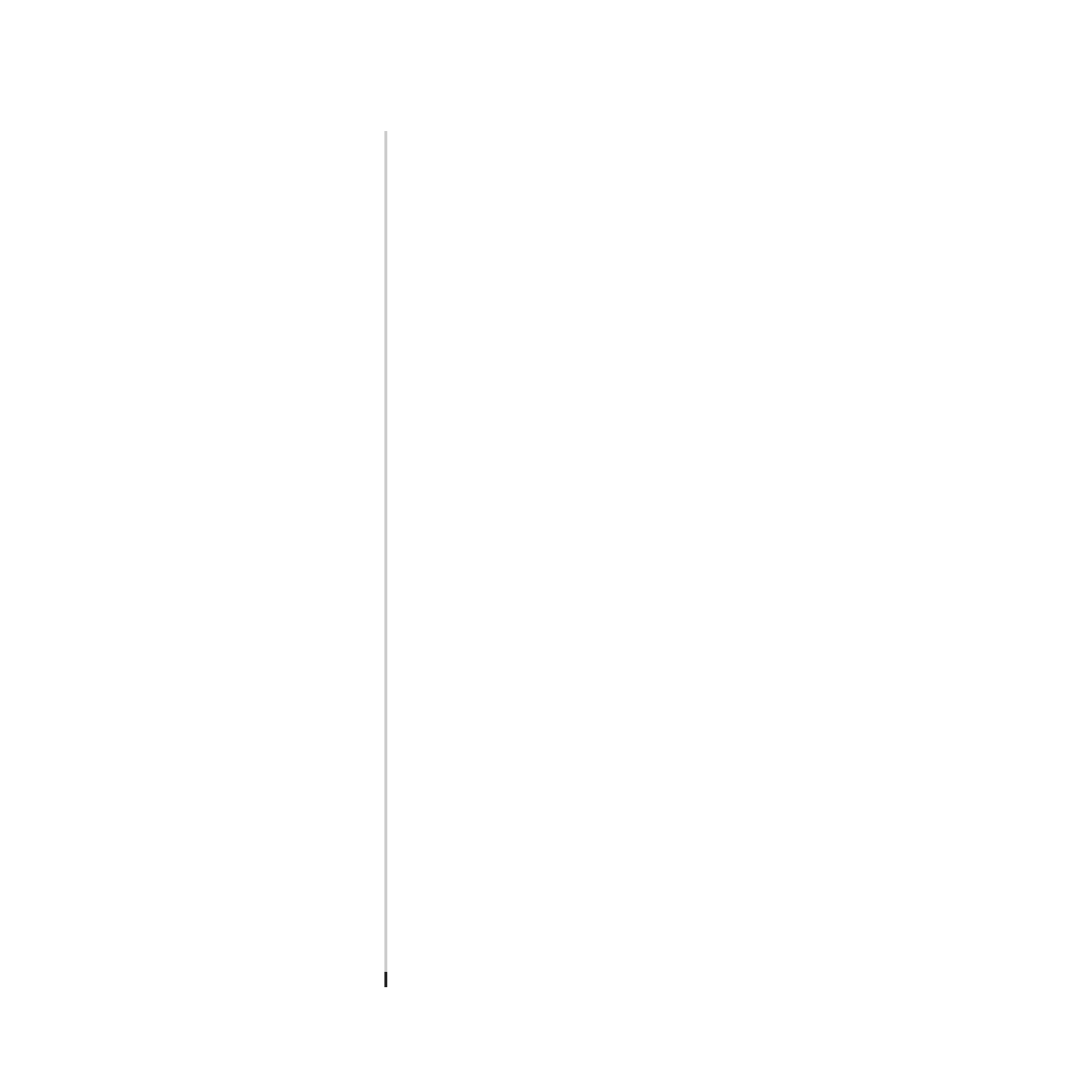}
         \caption{USS2}
         \label{fig:TSS3}
     \end{subfigure}
     \hfill
     \begin{subfigure}[b]{0.155\textwidth}
         \centerline{\def\svgwidth{80bp}
         \fontsize{7pt}{9pt}\selectfont
         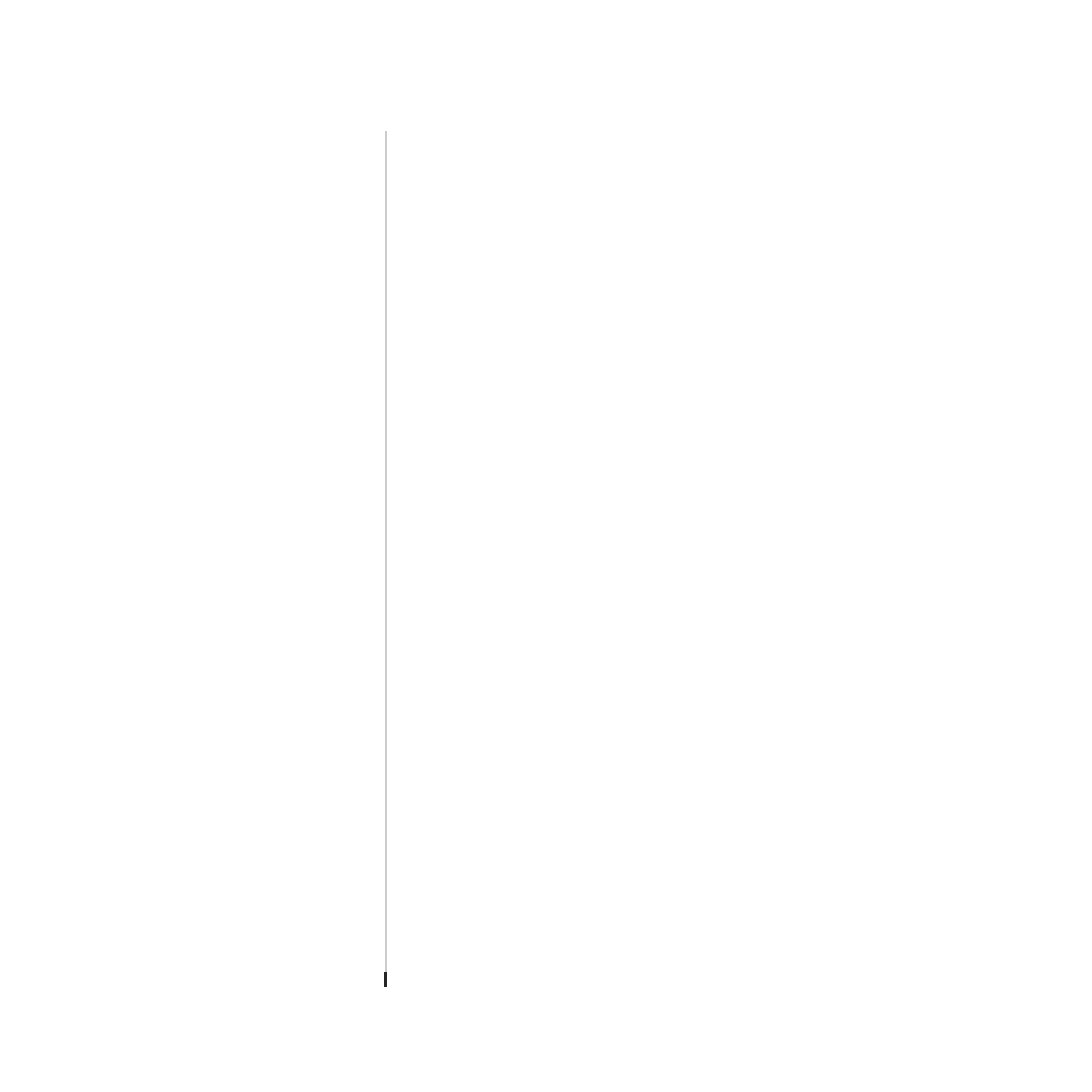}
         \caption{Unaltered Model}
         \label{fig:TSS3}
     \end{subfigure}
        \caption{Flexibility areas for $6$ scenarios with similar observable network component values but different unobservable network component values. Feasible flexibility samples (\protect\includegraphics[height=0.5em]{Figures/Colors/Orange.png}), infeasible flexibility samples (\protect\includegraphics[height=0.5em]{Figures/Colors/Blue.png}), initial operating point (\protect\includegraphics[height=0.5em]{Figures/Colors/Red.png}), and convex hull area of feasible samples (\protect\includegraphics[height=0.5em]{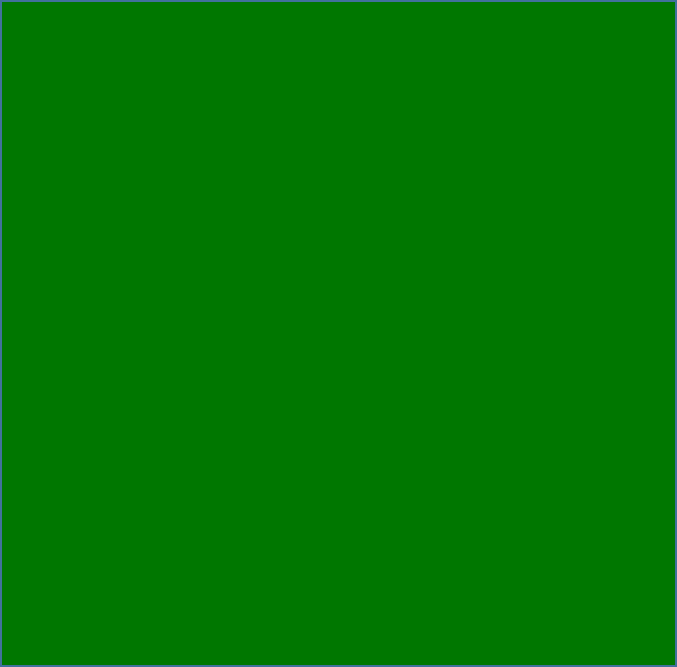}). }
        \label{fig:FLEX}
\vspace{-1em}
\end{figure}

\begin{table}
\caption{Convex hull area per scenario}
\label{tab:conv}
\begin{tabularx}{0.48\textwidth}{X|X|X}
\toprule
\multicolumn{1}{l|}{\textbf{Scenario}} & \textbf{Area} [$MW \cdot MVAR$] & \textbf{Area difference} [$\%$] \\
\midrule
\multicolumn{1}{l|}{TSS1} & 6.6 & -22.1 \\
\multicolumn{1}{l|}{TSS2} & 8.3 &	-1.5 \\
\multicolumn{1}{l|}{TSS3} & 12.1 & 42.7 \\
\multicolumn{1}{l|}{USS1} & 7.8 & -8.0 \\
\multicolumn{1}{l|}{USS2} & 6.9 & -18.9 \\
\multicolumn{1}{l|}{Unaltered Model} & 8.4 & 0.0\\
\bottomrule
\end{tabularx}
\vspace{-1em}
\end{table}

Concretely, Fig. \ref{fig:FLEX} depicts the flexibility area of each scenario. These areas are different from each other due to the aforementioned topological or operational differences. Table \ref{tab:ERR} displays the percentage difference between the altered $p_1,q_1$, and the initial $p_0,q_0$ network configurations' power flow for each observable line ($\frac{p_{1}-p_{0}}{p_{0}}$, $ \frac{q_{1}-q_{0}}{q_{0}}$). In contrast, Table \ref{tab:conv} exhibits the percentage difference of the convex hull areas between each scenario $cv_1$ and the unaltered model $cv_0$ ($ \frac{cv_{1}-cv_{0}}{cv_{0}}$). Table \ref{tab:ERR} illustrates less than $0.5\%$ difference in the observable lines power flow between each scenario. Therefore, by only observing the components of Table \ref{tab:ERR} and the output power of each FSP, any of the scenarios can be assumed representative of the true network state. However, Table \ref{tab:conv} illustrates that the flexibility areas of these scenarios can be $47.1\%$ different from the unaltered model's. To achieve these minimal differences in Table \ref{tab:ERR} and the vast differences in Table \ref{tab:conv}, the non-observable component, and topological shifts are chosen to keep the power from each feeder approximately the same, e.g., among other shifts in $USS1$ and $USS2$, a $0.55 [MW]$ reduction in load $L4$ is accompanied by a $0.55 [MW]$ increase in load $L14$.

In summary, the scenarios discretize and simplify \eqref{eq:sys1a} to assess steady-state performance. The complete network states $x_{sc}$ and voltages $v_{sc}$ per scenario differ, but the actions $u$ are similar for each scenario. Equation \eqref{eq:sys1b} returns approximately the same $\tilde{x}$ for these different scenario states $x_{sc}$, i.e., $\tilde{x} \approx r(x_{sc}, u, v_{sc},0)$ $\forall (x_{sc},v_{sc}) \in X_{sc}$, where $X_{sc}$ is the set of all scenarios' initial network states and voltages. Therefore, \eqref{eq:sys1c} takes approximately the same inputs for all scenarios. If the flexibility estimation objectives account for this issue by including $h \circ r(\cdot)$ as in \eqref{eq:range}--\eqref{eq:multi}, then the algorithms can be adapted to reduce the effect of low observability. However, the existing algorithms' objective of \eqref{eq:range2} only accounts for $h(\cdot)$ which cannot distinguish the flexibility areas of Fig. \ref{fig:FLEX} unless assuming higher network observability or $x_{sc}$ knowledge.

\section{Conclusion and future work}
The neglected low observability in DNs can significantly affect the accuracy of the estimated flexibility. This reduced accuracy can impact the SO applicability and adoption of existing flexibility estimation algorithms. This paper proposes generalizing and re-iterating the problem of DN flexibility estimation to account for the observability level. The proposed definition also introduces flexibility continuity and multiplicity as an additional objective of flexibility estimation algorithms to inform the SO and the market players. 

The future work includes developing algorithms that estimate flexibility using observable node data. Another focus of these data-driven algorithms will be the adaptability of estimated areas to new information, improved area computation speed, and multiplicity.

\section*{Acknowledgment}

This research is part of the research program 'MegaMind - Measuring, Gathering, Mining and Integrating Data for Self-management in the Edge of the Electricity System', (partly) financed by the Dutch Research Council (NWO) through the Perspectief program under number P19-25.

\bibliographystyle{IEEEtran}
\bibliography{bibliography}

\vspace{12pt}
\color{red}

\end{document}